\documentstyle[12pt]{article}
\newcommand{\al}{\alpha}
\newcommand{\Al}{\mbox{\boldmath $\alpha$}}
\newcommand{\ba}{\begin{array}}
\newcommand{\be}{\begin{equation}}

\newcommand{\bea}{\begin{eqnarray}}
\newcommand{\non}{\nonumber}
\newcommand{\ea}{\end{array}}
\newcommand{\ee}{\end{equation}}
\newcommand{\eea}{\end{eqnarray}}

\newcommand{\lm}{\lambda}

\newcommand{\om}{\omega}

\newcommand{\rf}[1]{(\ref{eq:#1})}
\newtheorem{th}{Theorem}[section]
\newtheorem{prop}[th]{Proposition}
\newtheorem{df}[th]{Definition}
\newtheorem{lem}[th]{Lemma}

\begin{document}
\title{Multivariable Invariants of Colored Links \\
Generalizing the Alexander Polynomial \footnote{Submmited to
the Proceedings of the Conference on Quantum Topology,
24-28 March 1993, Kansas State University, Manhattan, Kansas. }}
\author{Tetsuo Deguchi}
\date{}
\maketitle
\begin{center}
Department of Physics, University of Tokyo,  \\
Hongo 7-3-1, Bunkyo-ku, Tokyo 113, Japan \\
\end{center}

\abstract{We discuss multivariable invariants of colored links associated
with the $N$-dimensional  root of unity representation of the quantum group.
   The invariants for $N>2$ are generalizations of the multi-variable
	 Alexander polynomial. The invariants vanish for disconnected links.
	  We review the definition of the invariants through (1,1)-tangles.
When $(N,3)=1$ and $N$ is odd, the invariant does not vanish
for the parallel link (cable) of the knot $3_1$,
while the Alexander polynomial vanishes for the cable link. }

\newpage
\setcounter{equation}{0}
\renewcommand{\theequation}{1.\arabic{equation}}
\section{Introduction}

The multi-variable Alexander polynomial is an isotopy invariant
of colored links.  A variable is assigned on each component of colored links.
A hierarchy of multi-variable invariants of colored links which
generalize the multi-variable Alexander polynomial
was proposed through representations of the colored braid group. \cite{PRL}
In order to define link invariants, however,
we need a proper method for regularization of the Markov trace. \cite{clp}
The purpose of this paper is to review the definition
of the multi-variable invariants of colored links
from the viewpoint of the crossing symmetry, and
to show a systematic calculation of the invariants.
We see that the (multi-variable) invariants
are stronger than the Alexnader polynomial.

The new invariants are related to the quantum group $U_q(sl(2))$.
It was shown that the representation of the (colored) braid group
for the invariant \cite{PRL} is equivalent
to the R-matrix of an $N$-dimensional root of unity representation
of the quantum group. \cite{MPL} ( $N$ = 2, 3, $\cdots$.)
Hereafter we call the root of unity representation
nilpotent representation.

The main problem we encountered was the following.
The state sum for the nilpotent root of unity  representation
vanishes for any link diagrams.  The standard method for the Markov
trace gives a trivial Markov trace which vanishes for any braid,
and we have a trivial  invariant that vanishes for any knot.
For the case of the Alexander polynomial, however,
we can define a non-trivial invariants although
the standard state sum vanishes.

The state models for the Alexander polynomial have
also this vanishing property. \cite{Lee,graded,J.Murakami,KS1}
For the Alexander polynomial there are methods for regularizations
by using the Hecke algebra or the skein relation.
It seems, however,
that these methods do not hold
for the cases of the invariants
associated with the nilpotent representations.

This paper consists of the following.
In \S 2 we derive the R-matrix for the nilpotent representation.
In \S 3 we review the regularization given in \cite{clp}.
In \S 4 we give the Clebsch-Gordan coefficients (CGC)
of the nilpotent representation, and then we discuss
the crossing symmetry,  which is quite
different from the crossing symmetry of the standard
CGC of $U_q(su(2))$.  Using the crossing symmetry of the nilpotent CGC
we prove a proposition important for the
definition of the multi-variable invariants.
In \S 5 we calculate the multi-variable invariants
for the parallel links of 2-braid knots.

\setcounter{equation}{0}
\renewcommand{\theequation}{2.\arabic{equation}}
\section{ R matrix for nilpotent rep.}
We introduce  the defining relations of $U_q(sl(2))$
for the generators $e,f$, and $k$. \cite{Drinfeld,Jimbo1}
\be
kek^{-1} = q e, \quad kfk^{-1} =q^{-1} f, \quad
\mbox{[} e, f \mbox{]} = {\frac {k^{2}-k^{-2}} {q-q^{-1}} } .
\ee
The following formulas for the commultiplication $\Delta$,
the antipode $S$, and the counit $\epsilon$ on the generators
define the structure of the Hopf algebra
\bea
\Delta(k^{\pm1}) &=& k^{\pm1} \otimes k^{\pm1} , \quad
\Delta(e) = e \otimes k + k^{-1} \otimes e, \quad
\Delta(f) = f \otimes k + k^{-1} \otimes f, \non \\
S(k^{\pm1}) &=& k^{\mp1} , \quad S(e)=-q e, \quad S(f)=-q^{-1} f, \non \\
\epsilon(k^{\pm1})&=&1, \quad \epsilon(e)=\epsilon(f)=0 .
\eea
Hereafter we shall sometimes use the notation $k=q^{H/2}$.
We shall use the following $q$-analog notations:
\be
[n]_q  = {\frac {q^{n} - q^{-n}} {q - q^{-1}} } , \quad
[n]_q ! = \prod_{k=1}^{n} [k]_q , \quad
[p;n]_q ! = \prod_{k=0}^{n-1} [p-k]_q .
\ee
where $n$ is a positive integer and $p$ is a complex parameter. In particular,
we assume that $[0]_q!  =  [p;0]_q! = 1 $.
The universal $R$ matrix of $U_q(sl(2))$ ($q$ generic case)
is given by \cite{Drinfeld}
\be
{\cal R} = q^{H \otimes H /2} \sum_{n=0}^{\infty}
{\frac {q^{n(n-1)/2}} {[n]!} }
\left( (q-q^{-1}) q^{H/2} e \otimes f q^{-H/2} \right)^n .
\label{eq:universalR}
\ee

Let us consider the nilpotent root of unity representation.
\cite{Roche,Concini,Kac}.
Hereafter we assume that $q^{2N}=1$ and $q^2$ is
a primitive $N$-th root of unity.
When $N$ is even, $q$ is $2N$-th primitive root of unity ($q^N=-1$,
while when $N$ is odd, $q$ is $2N$-th ($q^N=-1$) or
$N$-th primitive root of unity ($q^N=1$).

We introduce $x=e^N, y=f^N$, and $k^{2N}$,
which  are central elements of $U_q(sl(2))$ at  $q$ roots of unity.
For the nilpotent root of unity representation
they are given by  $x=y=0$, and $z=q^{2Np}$, where $p$ is a complex parameter.
We note that for the cyclic representations  $x$,$y$, $z$ are given by
generic complex parameters.

We denote  by $V(p)$ the left $U_q(sl(2))$ module
for the nilpotent representation. The dimension of $V(p)$ is
given by $N$. The nilpotent representation has the following fusion rule.
\cite{Graph}
\be
V(p_1) \otimes V(p_2) = \sum_{p3} N^{p_3}_{p_1p_2} V(p_3),  \non
\ee
where
\bea
 N^{p_3}_{p_1p_2} = &1& \mbox{  if  }  p_3=p_1 +p_2 -n , (n=0,1, \cdots N-1),
\non \\
                    &0&  \mbox{ otherwise. }  \label{eq:fusion}
\eea
The fusion rule is quite different from that \cite{Jimbo1} for the
finite dimensional representations of $U_q(sl(2))$.
 We note that the fusion rule for the  semi-periodic representation is given
in \cite{Arn},
where $x = 0, y \ne 0$ and $z \ne 1$ or $x \ne 0, y = 0$ and $z \ne 1$.

We now consider  a matrix representation $\pi_p$
induced from the module $V(p)$.
Let   $e_0, \cdots, e_{N-1}$ be the basis of the module $V(p)$.
For $a,b =0,1, \cdots N-1$, we assume the following matrix representations
\cite{Roche}
\bea
\pi_p(e)_{ab} &=& \left( [2p - a] [a+1] \right)^{1/2} \delta_{a+1,b} , \non \\
\pi_p(f)_{ab} &=& \left([2p-a+1] [a] \right)^{1/2} \delta_{a-1,b} , \non \\
\pi_p(k)_{ab} &=& (q^{H/2})_{ab} =q^{p - a} \cdot \delta_{a, b} .
\label{eq:rep}
\eea
Putting the representations of the generators \rf{rep} into the universal
R-matrix \rf{universalR},
we have the matrix representation of the R-matrix
\bea
R^{p_1p_2}(+)^{a_1 a_2}_{b_1b_2} &=&
\left( (\pi_{p_1} \otimes \pi_{p_2}) {\cal R} \right)^{a_1a_2}_{b_1b_2}
\non \\
& = & \delta_{a_1 b_1-n}
\delta_{a_2, b_2+n}
q^{2(p_1-b_1)(p_2-b_2)-n(p_1-b_1) +n(p_2-b_2) -(n^2-n)/2} \times   \non \\
& \times & {\frac {(q-q^{-1})^n} {[n]!} }
{\sqrt{ [b_1;n]! [a_2;n]! [2p_1-a_1;n]! [2p_2-b_2;n]! }}  \quad .
\non \\
\label{eq:nilpotentR}
\eea
It is shown \cite{MPL}
that the representation of the colored braid group \cite{PRL}
is equivalent to the R-matrix \rf{nilpotentR}
for the nilpotent representation through some gauge transformations.

\setcounter{equation}{0}
\renewcommand{\theequation}{3.\arabic{equation}}
\section{Regularization}

We now discuss the construction of the colored link invariants \cite{clp}.
For oriented tangles we follow the notation of \cite{Turaev2} and \cite{clp}.

It is straightforward to construct invariants of oriented colored tangles
from the representations \cite{PRL} of the colored braid group. We can define
the tangle invariants by checking the Reidemeister moves \cite{Turaev2,Yetter}
for the tangle diagrams
 (see also \cite{clp}).
\footnote{For the oriented state model
the Reidemeister moves of  tangles
can be easily shown  by using the correspondence between
the Reidemeister moves and the basic relations of
exactly solvable models such as
the first, second inversion relations and
the Yang-Baxter relation. }
We denote by $\phi(T,\Al)$ the invariant for a colored oriented tangle
$(T,\Al)$,
where $\Al=(p_1,\cdots, p_n)$ and  $p_j$ denotes the color
of the component
$j$ of the oriented tangle $T$.

We first note the following property:
\begin{prop}
For any (0,0)-tangle $(T, \Al)$, the value
$\phi(T,\Al)$ vanishes.
\end{prop}

In order to have a nontrivial (multi-variable) invariant,
we introduce {\it a regularization procedure }.
 The regularization consists of the following
2 steps. \cite{clp}
(1) Choosing one component $s$ of the link $L$,
we define an invariant $\Phi(L,s,\Al)$.
(2) Dividing the invariant $\Phi(L,s,\Al)$ by a factor that depends on
$s$, we obtain
an invariant ${\hat \Phi}(L,\Al)$ for colored link $(L,\Al)$.

\par
(1) {\it 1-st step}

Let $T$ be a $(1,1)$-tangle.
We denote by $\hat T$ the link obtained
by closing the open string of $T$.
It is not difficult  to show the following proposition \cite{clp}.

\begin{prop}
Let $T_1$ and $T_2$ denote two $(1,1)$-tangles.
If $\hat T_1$ is isotopic to $\hat T_2$
as a tangle in $S^3$ by an isotopy
which carries the closing component of $\hat T_1$
to that of $\hat T_2$, then $T_1$ is isotopic to $T_2$ as a $(1,1)$-tangle.
\end{prop}

We give a comment on Prop. 3.2, which is useful for the regularization.
Let $T_1$, $T_2$ be the two (1,1)-tangles defined as in Prop. 3.2.
Thanks to Prop. 3.2,  we have  $\phi(T_1,\Al)=\phi(T_2,\Al)$.
We recall that $\phi$ is invariant
under the Reidemeister moves for tangles.  Thus
for a link $L$ we can define an invariant
by the value of the tangle invariant $\phi(T,\Al)$, where
$T$ is  a (1,1)-tangle and   ${\hat T}$ is equivalent to the link $L$.

Let us consider the value of $\phi$ precisely.
Let  $s$ denote the closing component (or edge) of $\hat T$.
We denote by $\phi(T,\Al)^{a}_b$
the value $\phi$ for the colored  tangle $(T,\Al)$
with variables $a$ and $b$
on the closing component (or, on the two edges of the closing component $s$).
Then we can show  \cite{clp} that
the value of  $\phi(T, \Al)^{a}_b$ does not depend on $a$ or $b$:
\be
\phi(T,\Al)^a_b = \lm \delta_{ab} .  \label{eq:irrep}
\ee
The property \rf{irrep} is derived from the irreducibility of
the nilpotent representation.

We now define an invariant $\Phi$.
Let  $(L,\Al)$ be a colored link and
$T$ be a (1,1)-tangle such that ${\hat T}=L$.
Let  $s$ denote the closing component of $T$.
Then we define $\Phi$ by $\Phi(L,s,\Al)$ = $\phi(T, \al)^0_0$.
Recall that  $\Phi$ is well-defined, i.e.
$\Phi(L,s,\Al)$ does not depend on a choice of $T$.

\par
(2) {\it 2-nd step}

We note that the value $\Phi(L,s,\Al)$ depends on the cutting component $s$.
The invariant $\Phi(L,s,\Al)$
is an isotopy invariant of colored link $L$ with
a particular choice of the component $s$ of the link $L$.
We now construct an isotopy invariant which does not
depend on $s$. In order to give a proper regularization,
we use the following proposition.
\begin{prop} [{\cite{clp}}]
For a link $L$ and its color $\Al = (p_1, \cdots, p_n)$,
we have
\be
 \Phi(L,s,\Al)
([2p_s; N-1]!)^{-1}
=\Phi(L,r,\Al)
([2p_{r}; N-1]!)^{-1} .
\ee
\end{prop}
The proof of this proposition has been given in Appendix C of \cite{clp}.
We shall show Prop. 3.3  by using the crossing
symmetry of CGC in \S 4.

Thus we arrive at new invariants
${\hat \Phi}(L, \Al)$ of colored oriented links $(L, \Al)$.
\begin{df} [{\cite{clp}}]
For a colored oriented link $(L,\Al)$,
we define an isotopy invariant $\hat \Phi $ of $(L,\Al)$ by
\be
\hat \Phi(L,\Al) = \Phi (L,s,\Al)
([2p_s; N-1]!)^{-1} .
\ee
\end{df}

The new colored link invariants are related to the
multi-variable Alexander polynomial.
It was shown by J. Murakami \cite{J.Murakami} that
a colored link invariant which corresponds to
$\hat \Phi(L, \Al)$ for the $N$=2 case
 a version of the multivariable Alexander polynomial
(the Conway potential function).
Therefore the new colored link invariants $\hat \Phi(L,\Al)$
for $N =3, 4,\cdots,$ are generalizations
of the multivariable Alexander polynomial.

We make a comment on the invariants for non-colored links.
We can consider invariants $\Phi(L,s,\Al)$
also for non-colored links $(L,\Al)$ for which  all the colors of the
components are equal: $p_1= \cdots = p_n=p$ ($\Al=(p,\cdots, p)$).
We emphasize the following fact.
It is not trivial to show that when $(L,\Al)$ is a non-colored link
with $n \ge 2$ ($\Al=(p,\cdots, p)$) the invariant $\Phi(L,s,\Al)$
is independent of $s$, i.e., independent of the choice of
the cutting component. Thanks to Prop. 3.3 , however,  we can show that
$\Phi(L,s,\Al)$ is independent of $s$  when $(L,\Al)$ is non-colored
($\Al=(p,\cdots,p)$).
Thus we can define a link invariant $\Phi(L)$ by
\be
\Phi(L) = \Phi(L,1,\Al),  \label{eq:linkinv}
\ee
where $L$ is a link,  and $\Al=(p_1, \cdots, p_n)$
with $p_1 = \cdots =p_n =p$.

Let us  consider  the one component case $n=1$.
 We consider a knot $K$.
We can define a knot invariant
$\Phi$ by making use of  Prop. 3.2
and the property \rf{irrep}.  We define the invariant $\Phi(K)$ by
\be
\Phi(K)=\Phi(K,1,\Al).
\ee

\setcounter{equation}{0}
\renewcommand{\theequation}{4.\arabic{equation}}
\section{CGC and crossing symmetry}
Let us discuss the Clebsch-Gordan coefficients of
the nilpotent representations of
$U_q(sl(2))$ \cite{Graph}.
We note that the CGC of finite dimensional representations
of $U_q(su(2))$ are given in \cite{kr}.

Let us consider the left $U_q(sl(2))$-module $V(p)$ in \S2.
Let the symbol $|p,z>$ $(z=0,1, \cdots, N-1)$ be the basis vector
of $V(p)$. \cite{Rose,Graph}
The actions of the generators on the basis of the module
are given by
\bea
e|p,z> & = & \sqrt{[2p-z+1][z]} |p,z-1> , \quad
f|p,z> = \sqrt{[2p-z][z]} |p,z+1> , \non \\
k|p,z> & = & q^{p-z} |p,z> . \label{eq:mod}
\eea
We define an inner product among the vectors in the module $V(p)$  by
\be
(|p,z_1>, |p,z_2>) =  \delta_{z_1,z_2} . \label{eq:inner}
\ee
We assume that two vectors belonging to modules with
different values of $p$ are orthogonal:
\be
(|p_1,z_1>, |p_2,z_2>) = 0, \mbox{ if } p_1 \ne p_2. \label{eq:inner2}
\ee
For the real forms of $U_q(sl(2))$
the inner products
were discussed in \cite{Liskova}.

We consider the fusion rule \rf{fusion}
in terms of the modules. Let $V(p_3)$ be one of the modules
obtained from the decomposition of the tensor product $V(p_1)\otimes V(p_2)$.
We note that $p_3$ is given by $p_3=p_1+p_2-n$, where
$n$ is an integer with the condition $0 \le n \le N-1$.
Let $|p_1,p_2; p_3,z_3>$ be the basis vector of the module $V(p_3)$ with the
value $z$ in \rf{mod} is given by $z_3$.
Let us  introduce some symbols.
Let $m,n,z,w$ be integers with $0 \le m,n,z,w \le N-1$.  We recall that
$z_1,z_2,z_3$ are integers satisfying
$0 \le z_1,z_2,z_3 \le N-1$.
Let the symbol $p(n)$ denote
$p(n)=p_1+p_2-n$.
We define the Clebsch-Gordan coefficients for the nilpotent
representations by the relation
\be
|p_1,p_2; p_3,z_3> = \sum_{z_1,z_2} C(p_1,p_2,p_3;z_1,z_2,z_3)
|p_1,z_1> \otimes |p_2,z_2>,
\ee
and the normalization condition
\bea
& &\sum_{z_1=0}^{min(n+z,n+w)}
C(p_1,p_2,p(n); z_1,n+z-z_1,z) \times \non \\
&\times & C(p_1,p_2,p(m) ; z_1, n+w-z_1,w+n-m) = \delta_{nm} \delta_{zw}.
\non \\
\label{eq:norm1}
\eea
An explicit expression  \cite{Graph} of the nilpotent CGC is given by
\bea
&& C(p_1, p_2, p_3; z_1, z_2, z_3)
= \delta(z_3, z_1+z_2-n) \times \non \\
&& \quad \times {\sqrt{[2p_1+2p_2-2n+1]}}
{\sqrt{[n]! [z_1]!  [z_2]!  [z_3]!}} \times
\non \\
&& \times \sum_{\nu=\mbox{max}(0, n-z_2)} ^{\mbox{min}(n, z_1)}
 {\frac {(-1)^{\nu} q^{-\nu(p_1+p_2+p_3+1)}
\times q^{(n-n^2)/2 + (n-z_2)p_1 +(n+z_1)p_2} }
{[\nu]! [n-\nu]! [z_1-\nu]!
[z_2-n+\nu]!}} \times \non \\
&& \times {\sqrt{\frac
{[2p_1-n; z_1-\nu] ! [2p_1-z_1; n-\nu]!
 [2p_2-n; z_2+\nu -n] ! [2p_2-z_2; \nu] ! }
 {[2p_1 +2p_2 -n+1; z_1+z_2+1]! }}} . \non \\
\label{eq:eppp}
\eea
We  note that the sum over  the integer $\nu$
in \rf{eppp} is taken  under the condition:
max $\{0, n-z_2 \}$  $\le \nu  \le$  min $\{n, z_1 \}$.
A derivation of the expression \rf{eppp} is shown in \cite{Graph}.
However, we shall give a simple derivation of the nilpotent CGC
in Appendix A.

Let us consider dual representations. We introduce some symbols for them.
Let $(\pi, V)$ be a set of a representation and its module.
Let $V^*=Hom(V,{\bf C})$
be the dual vector space of $V$. We denote $v^*(w)$ by $<v^*,w>$, where
$v^* \in V^*$ and $w \in V$.
The transposed representation $\pi^t$ of $\pi$ is defined by
$< \pi^t v^*, w> = <v^*, \pi w>$.
We now consider the nilpotent representation $\pi_p$ given in \rf{rep}
and the module $V(p)$.
We define the dual representation $\pi^{*}_p$ of $\pi_p$ by
\be
\pi^{*}_p(a) = \pi_p^t \mbox{ o } S(a) = \pi^t_p (S(a)), \quad
 \mbox{ for all } a \in U_q(sl(2)).
\ee
We now show that  the dual representation is
equivalent to a nilpotent representation.  Let us define
 ${\bar p}$ and ${\bar z}$ by
\be
 {\bar p}=N-1-p, \quad {\bar z}=N-1-z .
\ee
We call ${\bar p}$  and ${\bar z}$ conjugates of $p$  and $z$,
respectively.
Let us  introduce  an operator $w$ $\in$ $Hom(V(p),V^*(p))$.
We define $w$ by the following matrix elements
on the basis vectors of $V(p)$ and $V(p)^*$.
\be
(w)_{ab}=(-q)^{-b}q^{-(N-1)p}\delta_{{\bar a} b} ,
\quad \mbox{ for } a, b = 0, \cdots, N-1.
\ee
Then we see that $\pi^*_p$ corresponds to $\pi_{\bar p}$ through the
following relation
\be
\pi^*_p(a)  = w \pi_{\bar p}(a) w^{-1} , \quad
\mbox{ for all } a \in U_q(sl(2)).  \label{eq:dual}
\ee
We can show \rf{dual} by checking for the relations for the generators
$\pi_p^t(e)=\pi_p(f)$, $\pi^t_p(f)=\pi_p(e)$, $\pi_p^t(k)=\pi_p(k)$
and
\be
\pi_{\bar p}(e)_{{\bar a}{\bar b}} = \pi_p(f)_{ab}, \quad
\pi_{\bar p}(f)_{{\bar a}{\bar b}}=\pi_p(e)_{ab}, \quad
\pi_{\bar p}(k)_{{\bar a}{\bar b}}  = \pi_p(k^{-1})_{ab}.
\ee
We note that in the case of the finite dim. rep. of $U_q(su(2))$  \cite{kr} the
dual representation of a finite dim. rep. $j$ is equivalent to the rep.
$j$ itself.

We shall consider the  crossing symmetry  of the nilpotent CGC.
Hereafter we employ the following expression of CGC.
(We fix the phase factor related to the square root so that
the CGCs satisfy the simplest form of the crossing symmetry. )
\bea
&& C(p_1, p_2, p_3; z_1, z_2, z_3)
= \delta(z_3, z_1+z_2-n) \times \non \\
& & \qquad \times \sum_{\nu=\mbox{max}(0,n-z_2)}^{\mbox{min}(n,z_1)}
{\frac {(-1)^{\nu} q^{-\nu(p_1+p_2+p_3+1)} q^{(n-z_2)p_1 +(n+z_1)p_2} }
{ [\nu]! [n-\nu]! [z_1-\nu]! [z_2-n+\nu]!} } \times \non \\
&& \times \bigg( [2p_3+1][n]! [z_1]!  [z_2]!  [z_3]! q^{n-n^2}
\times  \non \\
&& \times {\frac {[2p_1-n; z_1-\nu] ! [2p_1-z_1; n-\nu]!
 [2p_2-n; z_2+\nu -n] ! [2p_2-z_2; \nu] ! }
 {[2p_1 +2p_2 -n+1; z_1+z_2+1]! }} \bigg)^{1/2} . \non \\
\label{eq:cgc}
\eea
Here  we note that the sum over $\nu$  is taken
under the condition:
max $\{0, n-z_2 \}$  $\le \nu  \le$  min $\{n, z_1 \}$.
Now we show the crossing symmetry of the nilpotent CGCs
\rf{cgc} which is consistent with the correspondence \rf{dual}.
\begin{prop} Crossing symmetry
\bea
C(p_3,{\bar p_2},p_1; z_3,{\bar z_2},z_1) & = &
(-1)^{z_3-z_1} q^{(p_2-z_2)-Np_2}
{\sqrt{\frac {[2p_3; N-1]!} {[2p_1; N-1]!}}} \times \non \\
&& \times C(p_1,p_2,p_3; z_1,z_2,z_3) ,
\label{eq:crossing}
\eea
where the symbols ${\bar p_2}$ and ${\bar z_2}$
denote the conjugates of $p_2$ and $z_2$, respectively,
and are given by
\be
{\bar p_2} = N-1 - p_2, \quad {\bar z_2} = N-1 -z_2.
\ee
\end{prop}
(Proof)
We shall give a proof of \rf{crossing} through induction on $z_1$
in Appendix B.

The crossing symmetry of the nilpotent CGC is
different from that of the CGC of the finite dim. representations
\cite{kr}.
We note that the crossing symmetry of the nilpotent CGC is
consistent with the crossing symmetry of  the
colored braid matrix (the nilpotent R matrix) given in
\S 6 of \cite{clp}. We also note that the latter
 can be derived from the relation
$(S \otimes 1) {\cal R} = {\cal R }^{-1}$ \cite{Drinfeld}
by using the correspondence \rf{dual}.

Making use of the crossing symmetry of the nilpotent CGC
we prove the next proposition.
\begin{prop} \cite{DO}
\bea
 \sum_{z_2=0}^{N-1} \left( C(p_1,p_2,p_3; z_1,z_2,z_1+z_2-n)
\right)^2 q^{2 \rho(p_2,z_2)}
 & = &  {\frac {[2p_3+1] [2p_1;N-1]!} {[2p_1+2p_2+1;N]!}}, \non \\
 \sum_{z_1=0}^{N-1} \left( C(p_1,p_2,p_3; z_1,z_2,z_1+z_2-n)
\right)^2 q^{- 2 \rho(p_1,z_1)}
& = & {\frac {[2p_3+1] [2p_2;N-1]!} {[2p_1+2p_2+1;N]!}},  \non
\eea
where $p_3=p_1+p+2-n$ and $\rho(p,z)$ is given by
\be
\rho(p,z) = (p-z) -Np .           \label{eq:prop-do}
\ee
\end{prop}
(Proof)  Using the crossing symmetry \rf{crossing}
and the orthonormal condition \rf{norm1} of the nilpotent CGC
we have the following.
\bea
& &  \sum_{z_2} \left( C(p_1,p_2,p_3; z_1,z_2,z_1+z_2-n)
\right)^2 q^{2 \rho(p_2,z_2)} \non \\
& =&  \sum_{z_2} \bigg( C(p_3,{\bar p}_2,p_1; z_1-{\bar z}_2 + {\bar n},{\bar
z}_2,z_1) \times \non \\
& \times & (-1)^{(z_3-z_1)}q^{-(p_2-z_2)+Np}
{\sqrt{\frac {[2p_1; N-1]!} {[2p_3; N-1]!}}} \quad
\bigg)^2 q^{2 \rho(p_2,z_2)}  \non \\
& = & {\frac {[2p_1; N-1]!} {[2p_3; N-1]!}} \sum_{{\bar z}_2}
\left( C(p_3,{\bar p}_2,p_1; z_1-{\bar z}_2 + {\bar n},{\bar z}_2,z_1)
\right)^2 \non \\
& = & {\frac {[2p_1; N-1]!} {[2p_3; N-1]!}}  \non \\
&= & {\frac {[2p_1; N-1]! [2p_3+1]} {[2p_1+ 2p_2 +1; N]!}}.
\eea

Let us discuss Prop. 3.3 in \S 3.
It is straightforward
 to show Prop. 3.3 from  Prop. 4.2. Thus from the crossing symmetry
we can derive Prop. 3.3 that is important in the definition of
the multi-variable invariants.

\setcounter{equation}{0}
\renewcommand{\theequation}{5.\arabic{equation}}
\section{Evaluation for cables of 2-braid knots}

We now discuss  calculation of   the invariant  $\Phi(L)$
for the parallel link of a 2-braid knot $b_1^k$.
 It is known that the Alexander polynomial vanishes
for the parallel link (the cable) of any knot.
Let us consider the parallel link $L$ of the trefoil knot.
We note that the trefoil knot is equivalent to
the closed braid of $b_1^3$.
The cable link $L$ gives an example that it is not a split link
but it has split Seifert surfaces,
and therefore the Alexander polynomial vanishes for it.
 \footnote{Prof. H.R. Morton suggested that the author should calculate
  the cable of the trefoil knot. \cite{Morton}}

Let us introduce a technique for evaluation.
We consider spectral
decomposition of the R matrix in terms of the nilpotent CGC.
We denote $C(p_1,p_2,p_1+p_2-n; a_1,a_2,a_1+a_2-n)$ by  $C_{12}(n; a_1,a_2)$.
We recall that $p_1,p_2$ $\in {\bf C}$  and $n=0,\cdots, N-1$.
We define  $\lambda_n(p_1,p_2)$ by
\be
\lambda_n(p_1,p_2) = (-1)^n q^{2p_1p_2 - 2n(p_1+p_2) +n(n-1)} .
\label{eq:eigen}
\ee
Then we have the following
\be
R^{p_1p_2}(\pm) = \sum_{n=0}^{N-1} (\lambda_n(p_1,p_2))^{\pm 1}
C_{12}(n; b_1,b_2)C_{21}(n; a_2,a_1). \label{eq:dec}
\ee
We give a comment on \rf{dec}.
The $\lambda_n(p_1,p_2)$ corresponds to the eigenvalue of the nilpotent
R-matrix. For the link polynomial associated with
the finite dimensional representation of $U_q(sl(2))$
\cite{Akutsu-Wadati,kr,Reshetikhin}
the eigenvalues of the R-matrix give the higher order skein relation.
 From the decomposition
\rf{dec} we can also derive "higher order" skein relation
for the multi-variable invariant.
For evaluation of the multi-variable invariants, however,
we shall use the decomposition \rf{dec} rather than the skein relation.
\footnote{We note that we can derive  \rf{dec} from
the spectral decomposition of the solvable vertex and IRF models associated
with the nilpotent representation  \cite{Graph}
, which we called colored vertex models and
colored IRF models.
The colored vertex and colored IRF models can be
constructed  by applying Jimbo's method in \cite{Jimbo1,Jimbo0}.}

Let us discuss the  normalization of the R-matrix.
We note  the following  relation \cite{clp} which corresponds to the
Reidemeister move I (or the Markov trace property)
\be
\sum_{b=0}^{N-1} R^{pp}(\pm)^{ba}_{ab} q^{2(p-b)-2Np} = q^{\mp2p{\bar p}} .
\label{eq:markov}
\ee
Let $f(p_1,p_2;q)$ be an  arbitrary function but satisfy the next condition:
\be
f(p_1 ,p_1; q) = q^{2 p_1 {\bar p}_1}  = q^{2 p_1 (N-1- p_1)}.
\ee
For example, we can consider
$f(p_1,p_2; q)= q^{p_1 {\bar p}_2 + {\bar p}_1 p_2}$.
We define  $\tilde R(\pm)$ by
\be
{\tilde R}^{p_1p_2}(\pm) = R^{p_1p_2}(\pm) (f(p_1,p_2; q))^{\pm 1} .
\ee
We use ${\tilde R}$ for calculation of
the (multi-variable) invariants. From \rf{markov} we see that
the normalization is consistent with  the Reidemeister move I (or the
Markov trace property).  \cite{clp}
We note that the property \rf{markov} was shown in \cite{clp} (see also
\cite{Zn-graded}).
We shall show another proof of \rf{markov} in Appendix C.

Using the decomposition \rf{dec} and Prop. 4.2  we now calculate $\Phi(K)$ for
2 braid knots $K$.
We assume that the knot $K$ is equivalent to the  closed braid of $b_1^k$.
(We also assume that $k$ is odd.)
Let the symbol $p_n$ denote $p_n=2p-n$. Then we have
\bea
\Phi(K)&=& \sum_{n=0}^{N-1} {\frac {[2p;N-1]!}{[2p_n;N-1]!} }
\left(\lambda_n(p,p)\right)^k q^{-k(2p^2-2(N-1)p)} \non \\
&=& {\frac {q^{2Np}-q^{-2Np}} {(q^{4Np}-q^{-4Np})(q^{2p+1}-q^{-2p-1})}}  \times
\non \\
&\times& \sum_{n=0}^{N-1} (-1)^n
q^{k(2(N-1)p-4np+n(n-1)}(q^{4p-2n+1}-q^{-4p+2n-1}) .
\label{eq:2knot}
\eea
 From the expression \rf{2knot} we see that
for $N>2$ we have $\Phi(b_1^k) \ne \Phi(b_1^{-k})$, in general.
We give, for example, the result for the trefoil knot $(k=3)$.
We use the notation that $\omega = q^{-2}, Z=q^{-4p}$. Then  we have
\bea
\Phi(b_1^3) = Z^{-2} + \om Z^{-1} + \om^2 - \om + Z + \om Z^2 , \non \\
\Phi(b_1^{-3}) = Z^2 + \om^2 Z + \om - \om^2 + Z^{-1} + \om^2 Z^{-2}.
\eea

For an illustration of the normalization, we show a calculation of
${\hat \Phi}(L, \Al)$, where $L$ is equivalent to the closed
(colored) braid  of $b_1^k$ with
$k$ even, and $\Al =(p_1,p_2)$.
We cut the first component with color $p_1$
and consider $\Phi(L,1, \Al)$. We recall that
$p(n)=p_1+p_2-n$. Then we have
\be
{\Phi}(L,1,\Al) = {\frac  {[2p_1; N-1]!} {[2p_1+2p_2+1; N]!} }
(f(p_1,p_2; q))^k
\sum_{n=0}^{N-1} [2p(n)+1] (\lambda_n(p_1p_2;q))^k .
\ee
We now divide
$\Phi(L,1,\Al)$ by $[2p_1; N-1]!$, and obtain ${\hat \Phi}(L,\Al)$:
\be
{\hat \Phi}(L,\Al) = {\frac  {(f(p_1,p_2; q))^k} {[2p_1+2p_2+1; N]!} }
\sum_{n=0}^{N-1} [2p(n)+1] \left(\lambda_n(p_1p_2;q)\right)^k .
\ee
Let $Z_{12}$ denote $Z_{12}=q^{2(p_1+p_2)} $.
If we choose the normalization
as $f(p_1,p_2; q) = q^{p_1 {\bar p}_2 + {\bar p}_1 p_2}$, then we have
\bea
{\hat \Phi}(L,\Al) & = & {\frac {(q-q^{-1})^N Z_{12}^{(N-1)k/2} }
{(Z_{12}^N-Z_{12}^{-N}) q^{N(N+1)/2} }} \times \non \\
&\times &
\sum_{n=0}^{N-1} \left( Z_{12}^{-kn+1}q^{kn(n-1)-2n+1}
-Z_{12}^{kn-1} q^{kn(n-1)+2n-1} \right).
\eea

Let us consider evaluation for the parallel links. We assume that
the link $L(k)$ corresponds
to the closed braid of $(b_2 b_1 b_3 b_2)^k b_3^{-2k}$ with $k$ odd, which
is the parallel link of the 2 braid knot $b_1^k$.
We again make use of Prop. 4.2 and the  decomposition \rf{dec}.
We have the following.
\bea
\Phi(L) & = & {\frac {[2p;N-1]!} {[8p+1;N]! (q-q^{-1})} } \times  \non \\
& \times & \sum_{m=0}^{N-1}(-1)^{mk}q^{-4k(2m-N+1)p+k(m^2-m)} \times \non \\
& \times & \left( q^{8p+1-2m} A_{N,m,k} - q^{-8p-1+2m} B_{N,m,k} \right) ,
\non
\eea
where
\bea
A_{N,m,k} & =&  N, \mbox{ if } (2m+1)k-2=0 \quad (mod N), \non \\
               & &  0, \quad \mbox{ otherwise }, \non \\
B_{N,m,k} & =&  N, \mbox{ if } (2m+1)k+2=0 \quad (mod N), \non \\
               & &  0, \quad \mbox{  otherwise }. \non \\
\label{eq:2braid}
\eea
{}From the evaluation \rf{2braid}
we obtain the following result.
\begin{prop}
The invariant for the parallel link  of 2-braid knot $b_1^k$
is non-zero if and only if $(k,N)$=1 and $N$ is odd.
\end{prop}

For an illustration we show the invariants for the parallel links
in some non-vanishing cases.
\bea
 \Phi(L(k))& = &\mp 5q^{-1}[2p;4]! (q-q^{-1})^4,  \non \\
&& \quad \mbox{ if } (k,N)=(3,5) \mbox{ and } q^5 = \pm1, \non  \\
\Phi(L(k)) & = &
\mp 7q^{-1}[2p;6]! (q-q^{-1})^6, \non \\
&& \quad
\mbox{ if } (k,N)=(3,7) \mbox{ and } q^7 = \pm1 , \non \\
\Phi(L(k)) & = & \pm 3q [2p;2]! (q-q^{-1})^2(q^{24p} + q^{-24p}), \non \\
&& \quad  \mbox{ if } (k,N)=(5,3) \mbox{ and } q^3 = \pm1 .
\eea

{}From the results of the
parallel links and the 2 braid knots  we can conclude
 that the (multi-variable)
invariants for the $N$ dim. nilpotent representations give
generalizations of the (multi-variable) Alexander polynomial.

\setcounter{equation}{0}
\renewcommand{\theequation}{6.\arabic{equation}}
\section{Discussion}
In the paper we have  calculated
the multi-variable invariants using the
Clebsch-Gordan coefficients of the nilpotent
representation. It is an interesting problem to evaluate
the multi-variable invariants for  closed
3 braids and their parallel links.
We can use  the expressions of the Clebsch-Gordan coefficients and the Racah
coefficients of the nilpotent representations obtained in  \cite{Graph}.
The results of the calculation  might suggest some unknown
 connections between    the knot theory \cite{Alexander,Conway} and
the invariants of the quantum group.

Let us consider the nilpotent
representation from
the viewpoint of cyclic representation of $U_q(sl(2))$ where
$x,y$, and $z$ are (generic) complex parameters.
The nilpotent representation could be viewed as  a special case of the
cyclic representation with parameters $x,y,z$.
However, the two representations give quite different R-matrices
\cite{Bazhanov,CVM} and different link invariants.
There are many works associated with  the cyclic, semi-periodic, and nilpotent
representations of $U_q(sl(n))$. \cite{Chak,Miki,Tarasov}
Precise connections among the R-matrices and link invariants
for the cyclic and the nilpotent representations should be discussed
elsewhere.

We now discuss the multi-variable
invariant from the viewpoint  of the Hopf algebras \cite{Drinfeld2}, and
 consider  the universal invariant related to the
multi-variable invariant of the knot case. \cite{Ohtsuki}
When $q$ is a root of unity,
the center of the universal enveloping algebra is larger than that for
the $q$ generic case, and  we can consider quotient algebras.
Reshetikhin and Turaev obtained a universal R-matrix for the
standard root of unity  representations of $U_q(sl(2))$, where
the generators $e$ and $f$ are nilpotent and $k^{2N}=id$.  \cite{Restur}
In the case of the nilpotent representation, the generators satisfy
relations $e^N=f^N=0$ and $k^{2N}=q^{2Np}, p \in {\bf C}$.
We may consider the defining relations of the
quotient algebra related to
 the nilpotent representation as a generalization of
those associated with  the standard root of unity representation.
Rosso obtained an expression of R matrices  for  the nilpotent
root of unity representations of $U_q(g)$. \cite{Rosso}
However, the R-matrices depend on the representations explicitly, and
are not operator-valued.
 A {\it colored ribbon Hopf algebra} is defined
in order to  employ the relation $k^N=q^{2Np}$
as one of the defining relations of the
quotient algebra,  and then a universal R-matrix for the
(colored) quotient algebra and
a universal invariant, whose value is given by an element of the
colored ribbon Hopf algebra,  are obtained. \cite{Ohtsuki}

Finally we give comments on the crossing symmetry of the nilpotent CGC.
The crossing symmetry of  the nilpotent CGC could be considered as
one of the most characteristic properties of  the
multi-variable invariants associated with the nilpotent representations.
There are several reasons. (1) We can consider Prop 3.3
as a consequence of the crossing symmetry of the nilpotent CGC.
We recall that Prop. 3.3 is important for the definition of
the multi-variable invariants.
(2) There are many different points in
the crossing symmetry of
the finite dimensional representations of $U_q(su(2))$
and that of the nilpotent representations.
(3) From the crossing symmetry of the nilpotent CGC
we can derive the crossing symmetries for the
colored braid matrix (the nilpotent R matrix) \cite{clp}
and for the lattice models \cite{CVM,Graph}.
The multi-variable invariants of colored framed graphs \cite{Graph}
also have the crossing symmetry.

{\vskip 1.2cm}
\par \noindent
{\bf Acknowledgements}
{\vskip 0.6cm}

The author would like to thank Y. Akutsu and T. Ohtsuki
for helpful discussions and their collaboration on the
multi-variable invariants.
He is  thankful to H.R. Morton for
helpful discussion both in his private communication
and in the Conference on  Quantum Topology, Kansas, March 24-28, 1993.
He is also thankful to K. Murasugi
for discussion on parallel links.
He thanks D. Yetter
and the other organizers of the Conference on Quantum Topology for their
kind invitation to the conference and hospitality there.
He is grateful to P.P. Martin for his kind invitation
to Dept. Math., the City University, where he completed the manuscript.
He would also like to thank M. Wadati for his keen interest in this work.

\setcounter{equation}{0}
\renewcommand{\theequation}{A.\arabic{equation}}
\renewcommand{\thesection}{A}
\section{Appendix A }
In \cite{Graph} an explicit proof for the
expression \rf{cgc} is given. Here, however, we give a simple
derivation of CGC \rf{cgc} using an operator notation.
We give the highest weight vector in the module $V(p_3)$
in an operator notation, which does not depend on an explicit matrix
representation. In Appendix A it is  not necessary to assume that $q$
is the root of unity, and the discussion holds also for the $q$ generic case.
\cite{Graph}

Let the $q$-combinatorial for a complex number $p$ and
a non-negative integer $z$ be
\be
\left[
\ba{c}
p \non \\
z
\ea
\right]   = {\frac {[p ; z]!} {[z]!}} .
\ee
Let $n$ be a non-negative integer.
We introduce a vector $[12;n,0>$ by
\bea
 \mbox{[}12; n,0> & = & \bigg(
\sum_{z_1=0}^{n} (-1)^{z_1} q^{-z_1(p_3+1)}
\left[
\ba{c}
2p_1-z_1 \non \\
n-z_1
\ea
\right] \times
\non \\
& \times &
\left[
\ba{c}
2p_2-(n-z_1) \non \\
z_1
\ea
\right]
 f^{z_1} \otimes f^{n-z_1} \bigg)
|p_1, 0> \otimes |p_2, 0> . \non \\
\label{eq:nn}
\eea
We shall show that $|p_1,p_2; p_3,0>$ is given by
$N(p_1,p_2,p_3) \mbox{[}12; n,0>$ with a normalization factor
$N(p_1,p_2,p_3)$.
We show the following.
\begin{lem} If we assume that $|p_i,0>$ are highest :
$e|p_i,0> = 0$ for $i=1$ and $2$, then  the vector
$\mbox{[}12;n,0>$ is also highest:
\be
\Delta(e) [n,0> = 0. \label{eq:highest}
\ee
\end{lem}
(Proof)
Let us define $z_2$ by $z_2=n-z_1$.
 We first calculate $[\Delta(e), f^{z_1} \otimes f^{z_2}]$.
\bea
&& [\Delta(e), f^{z_1} \otimes f^{z_2}]
= [e \otimes k,  f^{z_1} \otimes f^{z_2}] +
[k^{-1} \otimes e,  f^{z_1} \otimes f^{z_2}] \non \\
&=& [e,f^{z_1}] \otimes kf^{z_2}
+ (1-q^{z_2})f^{z_1} e \otimes kf^{z_2} \non \\
& + & k^{-1}f^{z_1} \otimes [e,f^{z_2}] + (1-q^{-z_1})k^{-1}f^{z_1}\otimes
f^{z_2} e .
\eea
Then we have
\bea
& & [\Delta(e), f^{z_1} \otimes f^{z_2}] |p_1,0> \otimes |p_2,0> \non \\
&=& \bigg({\frac {[z_1]} {q-q^{-1}}} q^{-z_2} f^{z_1-1}
(q^{1-z_1}k^2-q^{z_1-1}k^{-2}) \otimes f^{z_2}k \non \\
& + &  {\frac {[z_2]} {q-q^{-1}}} q^{z_1} f^{z_1} k^{-1}
\otimes f^{z_2-1} (q^{1-z_2}k^2-q^{z_2-1}k^{-2})\bigg) |p_1,0> \otimes |p_2,0>
\non \\
&=& \bigg([z_1]q^{p_2-z_2}[2p_1-z_1+1]f^{z_1-1} \otimes f^{z_2}
 + \non \\
& + & [z_2]q^{z_1-p_1}[2p_2-z_2+1]f^{z_1} \otimes f^{z_2-1} \bigg)
|p_1,0> \otimes |p_2,0>. \label{eq:com}
\eea
Substituting \rf{com} and \rf{nn} into $\Delta(e)\mbox{[}12; n,0>$ we have
the proposition.

We now assume the properties \rf{inner} and \rf{inner2} of the inner product
for $V(p_1)$ and $V(p_2)$.
Let us  assume the actions \rf{mod} of the generators on the base of
the modules $V(p_1)$ and $V(p_2)$, then we have
\be
|p_j,z_j> = f^{z_j} |p_j,0> / {\sqrt{[2p_j;z_j]! [z_j]!}} ,
\mbox{ for } j=1,2.  \label{eq:matrix}
\ee
We recall that $p(n)=p_1+p_2-n$, and that $z$ denotes a positive integer.
We now define  vector $\mbox{[}12;n,z>$  by
\be
\mbox{[}12;n,z> = (\Delta(f))^{z} \mbox{ [}12; n,0>
/ {\sqrt{[2p(n);z]! [z]!}} \quad . \label{eq:p3}
\ee
We can show that $[12;n,0>$
is orthogonal to $[12;m,z>$ if $m \ne n$. In particular,
we can show the following  by using \rf{matrix} (see \cite{Graph})
\be
([12; n,0>, [12; m,n-m>) = 0, \mbox{ if } 0 <  m \le n .
\ee
If we take the  normalization factor $N(p_1,p_2,p_3)$ as
\be
N(p_1,p_2,p_3) = q^{(n-n^2)/2 + np_2}
{\sqrt{\frac {[n]!} {[2p_3+n+1; n]! [2p_1; n]! [2p_2;n]!}}} ,
\ee
then we can show that the vector $N(p_1,p_2,p_3)[12;n,z>$
satisfies  the normalization condition of \rf{norm1}.
Thus we have shown that the vector  $[12; n,0>$ with
the normalization factor $N(p_1,p_2,p_3)$
 gives the normalized highest weight vector in the module $V(p_3)$
with $p_3=p(n)$.

We define $|p_1,p_2; p_3, 0>$ by
$|p_1,p_2; p_3, 0> = N(p_1,p_2,p_3)[12; n,0>$.
By applying $\Delta(f)$ to $|p_1,p_2; p_3, 0> = N(p_1,p_2; P_3)[12; n,0>$
and by using \rf{p3}
we obtain the Clebsch-Gordan coefficients of the nilpotent representation.

We make comments.
(1)  The lemma A.1  can be shown
 also for other representations with highest vectors such as
the infinite dim. representation, finite dim. rep. of $U_q(sl(2))$
with $q$ generic, and the semi-periodic rep. of $U_q(sl(2))$ at q roots of
unity. (2)
{}From the expression of the highest weight vector in
lemma  \rf{highest} we can also derive
the fusion rule and the Clebsch-Gordan coefficients
of the semi-periodic  representation. \cite{Talk}
Details will be discussed elsewhere.

\setcounter{equation}{0}
\renewcommand{\theequation}{B.\arabic{equation}}
\renewcommand{\thesection}{B}
\section{Appendix B}
We prove the crossing symmetry of the
CGC of the nilpotent representations by using induction on $z_1$.
\par \noindent
(1)For the case $z_1=0$,
 we can show the relation \rf{crossing} by
 explicit calculation using the expression \rf{cgc}.
We use the following properties of $q$-integers.
\bea
[N-p] & = & [p], \quad [p+1;N]! = (-1)[p; N]!, \mbox{ if } q^N=-1, \non \\
\mbox{[}N-p \mbox{]} & = &
(-1)[p], \quad [p+1; N]! = [p; N]! , \mbox{ if } q^N=1.
\eea

\par \noindent
(2) We assume that the relation
\rf{crossing} of the crossing symmetry holds for the case $z_1=z_1(0)$.
We consider  the recurrence  relations among CGCs.
We derive  the following two recursion relations
by considering  the actions of $\Delta(e)$ and  $\Delta(f)$
on the vectors of $V(p_3)$ for the
first and the second relations, respectively.
\bea
&& \sqrt{[z_3+1][2p_3-z_3]}C(p_1,p_2,p_3;z_1,z_2+1,z_3+1) =  \non \\
&& \sqrt{[z_1][2p_1+1-z_1]}q^{p_2-z_2-1}
C(p_1,p_2,p_3;z_1-1,z_2+1,z_3)  \non \\
& & + \sqrt{[z_2+1][2p_2-z_2]}q^{-p_1+z_1} C(p_1,p_2,p_3;z_1,z_2,z_3),
\label{eq:rec1} \\
&& \sqrt{[z_3][2p_3+1-z_3]}C(p_1,p_2,p_3;z_1,z_2-1,z_3-1) =  \non \\
&& \sqrt{[z_1+1][2p_1-z_1]}q^{p_2-z_2+1}
C(p_1,p_2,p_3;z_1+1,z_2-1,z_3)  \non \\
&& + \sqrt{[z_2][2p_2+1-z_2]}q^{-p_1+z_1} C(p_1,p_2,p_3;z_1,z_2,z_3).
\label{eq:rec2}
\eea
In the relation \rf{rec1}
we replace  $p_1,p_2,p_3, z_1,z_2$, and $z_3$ by
$p_3,{\bar p}_2, p_1, z_3+1,{\bar z}_2-1$, and $z_1(0)$, respectively,
and we have $(B.2)^*$.
We apply $(B.2)^*$ to the left hand side of the relation
\rf{crossing} for $z_1=z_1(0)$.  To the derived expression we apply
 the second relation \rf{rec2} with
 $z_2$ and $z_3$ replaced by $z_2+1$ and $z_3+1$, respectively.
 We then
obtain the crossing symmetry \rf{crossing} for the case $z_1=z_1(0)+1$.

\setcounter{equation}{0}
\renewcommand{\theequation}{C.\arabic{equation}}
\renewcommand{\thesection}{C}
\section{Appendix C}
We discuss the Markov trace property (the Reidemeister move I)
for the nilpotent R-matrix.
Using the decomposition \rf{dec} of the R-matrix
and Prop. 4.2 we show the following.
\begin{lem}
\be
\sum_{b=0}^{N-1} R^{pp}(\pm)^{ba}_{ab} q^{2(p-b)-2Np} = q^{\mp2p{\bar p}} .
\ee
\end{lem}
(Proof) We consider the case of + sign.
\bea
(LHS) & =& {\frac {[2p; N-1]!} {[4p+1; N]!}} \sum [4p+1-2n] \lambda_n(p,p;q)
\non \\
& =& {\frac {[2p; N-1]!} {[4p+1; N]!} } q^{-2p {\bar p}} (SUM)
{\frac 1 {q-q^{-1}}} , \non
\eea
where the $(SUM)$ is given by
\bea
(SUM) &=& \sum (-1)^n q^{2(N-1-2n)p + n(n-1)} (q^{4p+1-2n} -q^{-4p-1+2n})
\non \\
& = & \sum_{n=-1}^{N-2} - \sum_{n=1}^{N} (-1)^{n-1}
q^{2(N-1-2n)p+n(n-1)-1}  \non \\
& =& q^{2(N+1)p+1} -q^{2(N-1)p-1} - (-1)^{N}q^{-2(N-1)p+N(N-1)+1} + \non \\
& + & (-1)^N q^{-2(N+1)p+N(N-1)-1} .
\eea
We now note the following.
\be
[p;N]! = q^{(N-1)N/2} (q^{Np}-q^{-Np}) (q-q^{-1})^{-N}. \label{eq:formula}
\ee
Then we have
\bea
(LHS) & = & {\frac {q^{2Np}-q^{-2Np}}
{(q^{4Np+N} -q^{-4Np-N})(q^{2p-N+1}-q^{-2p+N-1}) }} (SUM) q^{2p^2-2(N-1)p}
\non \\
& = & q^{2p^2-2(N-1)p} .
\eea
The last equation can be shown by an explicit calculation using
$q^N = \pm 1$.
\end{document}